\documentstyle [12pt] {article}

\newcommand{\be}{\begin{equation}}
\newcommand{\ee}{\end{equation}}
\newcommand{\beqs}{\begin{eqnarray}}
\newcommand{\eeqs}{\end{eqnarray}}
\newcommand{\shalf}{\hbox{$\frac{1}{2}$}} 

\def\[{\left[}
\def\]{\right]}
\def\({\left(}
\def\){\right)}

\def\z{\zeta}
\def\d{\delta}

\def\a{\alpha}
\def\b{\beta}

\def\da{\dot{\alpha}}
\def\db{\dot{\beta}}
\def\pa{\partial}
\def\e{\epsilon}

\def\S{\Sigma}
\def\th{\theta}
\def\tp{\tilde{\Psi}}

\def\ni{\noindent}
\def\nn{\nonumber}
\def\tQ{{\cal Q}}
\def\tP{{\cal P}}
\def\hq{\hat{q}}
\def\hp{\hat{p}}
\def\hab{\hat{c}}
\def\hps{\hat{\psi}}

\def\ddbgs{(D\bar{D}G)^2}
\def\dbdgs{(\bar{D}DG)^2}
\def\ddbfs{(D \bar{D} \S)^2}
\def\dbdfs{(\bar{D} D \bar{\S})^2}
\def\s{\sigma}

\def\dblad{\bar{D}_{\da}}

\def\dla{D_{\alpha}}

\begin{document}

\begin{titlepage}

\begin{flushright}
\begin{tabular}{l} 
ITP-SB-98-24 \\
hep-th/9811130 \\ 
November, 1998
\end{tabular}
\end{flushright}

\vspace{8mm}
\begin{center} 

{\Large \bf On dual 3-brane actions with partially broken $N=2$ supersymmetry}

\vspace{20mm}

F.Gonzalez-Rey \footnote{email: glezrey@insti.physics.sunysb.edu},
I.Y. Park  \footnote{email: ipark@insti.physics.sunysb.edu} and         
M. Ro\v{c}ek \footnote{email: rocek@insti.physics.sunysb.edu},

\vspace{4mm} Institute for Theoretical Physics \\ 
State University of New York	\\ 
Stony Brook, N. Y. 11794-3840 \\

\vspace{20mm}

{\bf Abstract}
\end{center}
The $N=1$ superspace generalization of the 3-brane action in 6 dimensions
with partially broken $N=2$ supersymmetry can be constructed using 
$N=1$ chiral, complex linear, or real linear superfields. The physical 
scalars of these multiplets give equivalent descriptions of the two 
transverse coordinates. The second supersymmetry is realized nonlinearly 
in all these actions. We derive the superspace brane actions and their 
nonlinear supersymmetry for both kinds of linear superfields  
when we break $N=2$ supersymmetry spontaneously. This breaking is 
realized in the free action of hypermultiplets that live in $N=2$ 
projective superspace by constraining the $N=2$ multiplets to 
reduce them to a pure Goldstone multiplet. For the chiral 
superfield, the superspace brane action and nonlinear supersymmetry can 
be deduced by dualizing the brane action of either linear superfield. 
We find that the dual action is unique up to field redefinitions that 
introduce arbitrariness in the dependence on the auxiliary fields.

\vspace{35mm}

\end{titlepage}
\newpage
\setcounter{page}{1}
\pagestyle{plain}
\pagenumbering{arabic}
\renewcommand{\thefootnote}{\arabic{footnote}} \setcounter{footnote}{0}

\section{Introduction}

The manifestly $N=1$ supersymmetric form of 3-brane actions in six
dimensions \cite{Liu} with a
nonlinearly realized extended supersymmetry has been recently proposed
by different authors \cite{BG,RT}. The origin of this
additional symmetry can be traced to the spontaneous partial breaking
of a linearly realized $N=2$ supersymmetry \cite{BW} in the action of
hypermultiplets that live in projective superspace\footnote{For a review of
Projective superspace see \cite{proj} and references therein.}. A similar
partial supersymmetry breaking has also been studied in the 
$N=2$ super Maxwell action \cite{APT,BG1,IZ,RT}. Imposing the additional
constraint that the $N=1$ chiral superfields contained in these N=2
multiplets are nilpotent, the $N=1$ superspace action is uniquely
determined for the $N=1$ Maxwell and tensor multiplets \cite{RT}.

In this article, we
study the partial breaking of $N=2$ supersymmetry in the action of another 
off-shell description of the hypermultiplet (the $O(4)$ projective 
multiplet). This description involves a chiral, a complex linear\footnote{
The complex linear multiplet was first discussed in \cite{gates_siegel}. 
The on-shell description of a $N=2$ hypermultiplet involving a chiral 
and a complex linear multiplet was discussed in \cite{deo_gates}.} and a
real auxiliary $N=1$ superfield. Imposing the nilpotency constraint in
the chiral component we find the corresponding $N=1$ action of the
complex linear multiplet with extended supersymmetry.

 These actions with partially broken $N=2$ supersymmetry  
are nonlinear functionals of real or complex linear superfields in
the case of $O(2)$ and $O(4)$ hypermultiplets respectively. These
fields can be dualized by introducing chiral Lagrange multipliers that
impose the corresponding linearity constraint. Using the duality 
equations that relate the linear superfields and their chiral duals
we derive the nonlinear action of the latter. This action is
essentially the functional proposed in \cite{BG,RT} up to
terms depending on auxiliary fields that can be eliminated by a field
redefinition.

 We also compute the x-space components of the $N=1$ complex linear 
superfield action with partially broken supersymmetry. We find that 
they are of the same form as the bosonic action computed for the dual 
chiral superfield. This is consistent with the fact that in this case the
duality only involves the auxiliary components of the complex linear
superfield: its physical degrees of freedom are the same as those of
the chiral dual\footnote{ In contrast, the real linear superfield 
has a bosonic action that can be expressed in terms of dual
gauge tensor fields \cite{RT}.}.    

 Finally we study the effect of the superpotential deformation on the 
duality transformation of the $O(4)$ multiplet. We find that the dual
action is unchanged after rescaling the dual
chiral field by a constant that depends on the deformation parameter. 

\section{Partial breaking of $N=2$ supersymmetry in the hypermultiplet
 action}

 In this section we review the partial breaking of $N=2$ supersymmetry
in the free action of $O(2)$ hypermultiplets \cite{RT}. Then we derive
a similar partial breaking in the $O(4)$ case. This
last case can be straightforwardly generalized to the generic $O(2p)$ 
multiplet but the qualitative features of the symmetry breaking are
the same and therefore we will restrict our analysis to the $O(4)$
multiplet.   

 The $O(2p)$ hypermultiplet can be parameterized using a complex 
coordinate $\z$

\be 
 \eta = \sum_{n=-p}^{+p} \eta_n \z^n  \ .
\ee

\ni 
The component superfields obey the constraints\footnote{For notational 
simplicity we write the supercovariant derivatives of both 
supersymmetries as follows $D_{1 \a} = D_\a,\: D_{2 \a} = Q_\a$.}

\be
 Q_\a \, \eta_n = - D_\a \, \eta_{n+1} \;, \;\;\; \bar{Q}_{\da} \, \eta_n =
 \bar{D}_{\da} \, \eta_{n-1} \ , 
\label{proj_const}
\ee

\ni 
and these guarantee that any action of the type

\be
 S_{O(2p)} = (-)^p {1 \over 2} \int d^4 x \, D^2 \bar{D}^2 
 \oint {d \z \over 2 \pi i \z} F(\eta, \z)
\label{n2_act}
\ee

\ni 
is off-shell $N=2$ supersymmetric. In particular for the $O(2)$ multiplet

\be
\eta = { {\bf \bar{\Phi}} \over \z} + {\bf G} - {\bf \Phi} \z
\ee

\ni
the constraints imply $\bar{D}_{\da} {\bf \Phi} = 0 = D^2 {\bf G} =
\bar{D}^2 {\bf G}$, {\em i.e.}, $\eta$ contains an $N=1$ chiral 
scalar superfield, its conjugate, and an $N=1$ real linear superfield 
(also known as tensor
multiplet).  We explicitly break the second supersymmetry while
we preserve $N=1$ and Lorentz invariance by giving ${\bf \bar{\Phi}}$
and ${\bf G}$ a nonzero {\em v.e.v.}

\beqs
 {\bf \bar{\Phi}} & \equiv & \bar{\Phi} + (\theta^2)^2 \: , \;\;  
 \; \langle \bar{\Phi} \rangle =0    \nn   \\
 {\bf G} & \equiv & G - \th^{1 \a} \th^2_\a - 
 \bar{\th}_1^{\da} \bar{\th}_{2 \da} \; , \;\;\; \langle G \rangle = 0  \ .
\eeqs

\ni
The constraints (\ref{proj_const})
can be rewritten in terms of the shifted $N=1$ chiral and tensor 
superfields\footnote{We follow the superspace conventions of
\cite{book}. In particular $D_\a \th^\b = \d_\a^\b$. } 

\beqs
& & Q_\a \bar{\Phi} = - D_\a G \nn \\
& & Q_\a G = \th^1_\a + D_\a \Phi  \nn \\
& & Q^2 \bar{\Phi} - 1 = - D^2 \Phi  \ .
\eeqs

\ni 
If we impose the additional constraint $\Phi^2 = 0 = \bar{\Phi}^2$ we
find

\be
 Q^2 \({1 \over 2} \bar{\Phi}^2 \) = 0 \Longrightarrow 
 \bar{\Phi} = D^2 \( \Phi \bar{\Phi} - {1 \over 2} G^2 \) \Longrightarrow 
 \bar{\Phi} = - {1 \over 2} { (D^\a G)(D_\a G)
 \over 1 - D^2 \Phi }
\label{const_sol}
\ee

\ni
and the corresponding conjugate. This implies that any action of the form
$S_{O(2)}$ (\ref{n2_act}), including the free action$ \int d^4x \, 
d^4 \th \, ( \Phi \bar{\Phi} - \shalf G^2 )$, is proportional to 
$\int d^4 x \, d^2 \th \Phi(G)$. An illuminating form of the action is

\be
 S_{O(2)} =
 \int d^4 x \, D^2 \bar{D}^2 \( {1 \over 4} { D^\a G D_\a G \bar{D}^{\da} G
 \bar{D}_{\da} G  \over (1 - \bar{D}^2 \bar{\Phi}) (1 - D^2 \Phi) } - 
 {1 \over 2} G^2 \) \ .
\label{action_2}
\ee

 Since the first term in the action (\ref{action_2}) contains the 
maximum number of fermionic superfields $\Psi = D G, \bar{\Psi} = 
\bar{D} G$, we can replace $\bar{D}^2 \bar{\Phi}$ and $D^2 \Phi$ in 
the denominator by the solution of the system of equations

\beqs
 \bar{D}^2  \bar{\Phi} & = & {1 \over 2} { (\bar{D}^{\da} D^\a G)
 (\bar{D}_{\da} D_\a G) \over 1 - \bar{D}^2 \bar{\Phi} } + O(\Psi) \nn \\
 D^2 \Phi & = & {1 \over 2} { (D^\a \bar{D}^{\da} G) (D_\a \bar{D}_{\da} G) 
 \over 1 - \bar{D}^2 \bar{\Phi} } + O( \bar{\Psi} )  \ ,
\eeqs

\ni
where we have dropped terms linear or quadratic in the fermionic
fields. The solution to the resulting quadratic equation must be
chosen to avoid giving a vacuum expectation value to the bosonic
superfield $D^2 \phi$, as that would break $N=1$ supersymmetry. With such a
choice the resulting nonlinear action is \cite{BG,RT}

\be
 S_{G} = \int d^4 x \, d^4 \th \( - {1 \over 2} G^2 + 2 \Psi^2
 \,\bar{\Psi}^2 \,f \) \ .  
 \label{Gaction}
\ee

\ni 
We follow the notation in \cite{RT,BG}
 
\beqs
 \Psi^2 & \equiv & {1 \over 2} D^\a G D_\a G      \\ 
 f & \equiv & {1 \over 1 - A + \sqrt{1- 2 A + B^2 } }  \nn \\
 A & \equiv & {1 \over 2} \( \ddbgs + \dbdgs \) \nn \\
 B & \equiv & {1 \over 2} \( \ddbgs - \dbdgs \) \nn \ .     
\label{BGnot}
\eeqs

This action is by construction invariant under the nonlinear supersymmetry
transformations 

\be
 \d G = (\e^\a Q_\a + \bar{\e}^{\da} \bar{Q}_{\da} ) G = 
 \e^\a \( \th_\a + D_\a \Phi(G) \) + \bar{\e}^{\da} 
 \( \bar{\th}_{\da} + \bar{D}_{\da} \bar{\Phi}(G) \)   \ . 
\label{nonlin_transf}
\ee

 Now we turn our attention to other off-shell representations of the 
$N=2$ algebra describing the same degrees of freedom, and try to
implement a similar partial breaking of $N=2$ supersymmetry on their 
free action. The simplest of these representations is the projective 
$O(4)$ multiplet. The projective constraints (\ref{proj_const}) tell
us that this multiplet contains a $N=1$ chiral superfield and its
conjugate, a $N=1$ complex linear superfield 
\cite{gates_siegel, deo_gates} and its conjugate, and a real 
unconstrained superfield

\be
 \eta = {{\bf \bar{\Phi}} \over \z^2} + {{\bf \bar{\S}} \over \z} + 
 X - {\bf \S} \z + {\bf \Phi} \z^2 \; , \;\;\; D_\a {\bf \bar{\Phi}} 
 = 0 = D^2 {\bf \bar{\S}} \; , \;\;\; \bar{X} = X \ .
\ee

\ni 
A vacuum expectation value for the chiral field and the complex linear
field induces modified constraints as before

\beqs
 {\bf \bar{\Phi}} = \bar{\Phi} + (\th^2)^2 & \Longrightarrow & 
 Q_\a \bar{\Phi} = - D_\a \bar{\S} \nn \\
 {\bf \bar{\S}} = \bar{\S} - \th^{1 \a} \th^2_\a & \Longrightarrow & 
 Q_\a \bar{\S} = \th^1_\a - D_\a X  \nn \\
& & Q^2 \bar{\Phi} = 1 + D^2 X \ .
\eeqs

The additional constraint $\bar{\Phi}^2=0$ once again gives
a solution in terms of the fermionic superfields
$\tp_\a = D_\a \bar{\S}, \bar{\tp}_{\da} = \bar{D}_{\da} \S$ and the
auxiliary field $X$ 

\be
 \bar{\Phi} = - {1 \over 2} { \tp^\a \tp_\a \over 1 + D^2 X }
\label{const_soll}
\ee

\ni 
In this case we have the auxiliary 
superfield $X$ in the denominator and we cannot eliminate it as easily as
before. We may, however, substitute (\ref{const_soll})
into the free action of the $O(4)$ multiplet

\beqs
 S_{O(4)} & = & \int d^4 x \, d^4 \th \( \Phi \bar{\Phi} - \S \bar{\S} + 
 {1 \over 2} X^2 \)            \label{O4_nonl_act}        \\
 & = & \int d^4 x \, d^4 \th \(- \S \bar{\S} + {1 \over 4} 
 { \tp^\a  \tp_\a \bar{\tp}^{\da} \bar{\tp}_{\da}  
 \over (1 + D^2 X) (1 + \bar{D}^2 X) } + {1 \over 2} X^2 \) \nn
\eeqs

\ni
and then obtain the algebraic field equation of the auxiliary $X$

\be
 X = {- 1 \over 4 (1 + \bar{D}^2 X) (1 + D^2 X) } \( \tp^\a \tp_\a
 {\ddbfs \over 1 + D^2 X} + \bar{\tp}^{\da} \bar{\tp}_{\da} 
 { \dbdfs \over 1 + \bar{D}^2 X } \) + 
 O(\tp^2 \bar{\tp}, \tp \bar{\tp}^2) \ .
\ee

 We have dropped terms with more than two fermionic superfields because
they do not contribute to the term $X^2$ in the action. For the same
reason the term with four fermionic fields picks up the purely bosonic
part of $D^2 X$ and $\bar{D}^2 X$

\beqs 
 D^2 X & = & {1 \over 4} {\ddbfs \dbdfs \over (1 + D^2 X) 
 (1 + \bar{D}^2 X)^2 } + O(\tp, \bar{\tp} ) \nn \\
 \bar{D}^2 X & = & {1 \over 4} {\ddbfs \dbdfs \over (1 + D^2 X)^2 
 (1 + \bar{D}^2 X) } + O(\tp, \bar{\tp}) \ .
\eeqs

\ni
From this identity we learn that the purely bosonic part of this
superfield is real $D^2 X = \bar{D}^2 X$. For notational simplicity we
work from now on with the shifted real field 

\be
 Y= 1 + D^2 X \ ,
\ee

\ni 
which obeys 

\be
 Y^3(Y+1) = {1 \over 4} \ddbfs \dbdfs \ .
\label{implicit}
\ee

\ni
It is possible to solve this equation for $Y$ and choose the root
with unit {\em v.e.v.}

\be
 Y = {1 \over 4} + { {\cal V}^{1 \over 2} \over 12} +
 {\sqrt{6} \over 12} \sqrt{ \( 3 {\cal Z}^{1 \over 3} - {\cal Z}^{2 \over 3} + 
 12 \ddbfs \dbdfs \) {\cal V}^{1 \over 2} + 9 {\cal Z}^{1 \over 3}   
 \over {\cal Z}^{1 \over 3} {\cal V}^{1 \over 2} }
\ee

\ni
where

\beqs
 {\cal Z} & = & -27 \ddbfs \dbdfs + 3 \sqrt{ 81 \(\ddbfs \dbdfs \)^2 + 
 192 \(\ddbfs \dbdfs \)^3 } \nn  \\
 {\cal V} & = & 9 + 6 {\cal Z}^{1 \over 3} - 72 
  { \ddbfs \dbdfs \over {\cal Z}^{1 \over 3} } \ .
\eeqs

\ni
However, we find it more useful to maintain
the dependence of the action on the real superfield $Y$. The
resulting nonlinear action is

\be
 S_{O(4)} = \int d^4 x \, d^4 \th \( - \S \bar{\S} +  
 \tp^2 \bar{\tp}^2 {2 Y + 1 \over Y^3} \) \ ,
\label{Sigma_act}
\ee

\ni
and by construction it is also invariant under the nonlinear
supersymmetric transformations 

\be
 \d \S = (\e^\a Q_\a + \bar{\e}^{\da} \bar{Q}_{\da} ) \S = 
 \e^\a D_\a \Phi(\S, \bar{\S}) + \bar{\e}^{\da} 
 \(\bar{\th}_{\da} - \bar{D}_{\da} X(\S,\bar{\S}) \) \ . 
\ee

 It is interesting to note that in this case we can add a superpotential 
deformation proportional to the constrained chiral superfield, which 
is still $N=2$ supersymmetric but not the same as the nonlinear action

\beqs
 S_\Phi & = & {\b \over 2} \int d^4 x \, d^4 \th \oint {d \z \over 2 \pi i \z} \;
 \eta^2 \( \z^2 + {1 \over \z^2} \) = \b \int d^4 x D^2 \bar{D}^2 \( \Phi X + 
 {1 \over 2} \S \S + c.c. \)  \label{deform}  \\
& = & \b \int d^4 x D^2 \( \Phi \bar{D}^2 X + \bar{\tp}^2  \) + c.c. = \b
 \int d^4 x D^2 \( \Phi \bar{D}^2 X - \Phi (1 + \bar{D}^2 X) \) + c.c.  \nn \\
& = & - \b \int d^4 x D^2 \: \Phi \nn  \ .
\eeqs

In the last section we will prove that adding this superpotential  
deformation only amounts to a change in the normalization of the dual 
chiral Lagrange multiplier and the appearance of an overall factor
multiplying the dual action. Such rescaling is well defined for any 
value of the parameter $\b$ except $\b= \pm 1$ where it becomes 
singular\footnote{The pure free $N=1$ complex linear superfield action
including a quadratic term $\b (\S^2 + \bar{\S}^2)/2$, was first considered in 
\cite{deo_gates}, where it was observed that the value $\b= \pm 1$ is 
critical and needs special treatment.}.

\section{Dual nonlinear actions}

 It is well known that the $N=1$ supersymmetric actions of real linear 
and complex linear superfields can be dualized into actions of chiral
fields describing the same on-shell degrees of freedom. For the real
linear superfield this is a true T-duality that relates different physical
x-space fields of the corresponding nonlinear sigma models. For the 
complex linear superfield it is a pseudo-duality that merely changes
the auxiliary fields without affecting the physical ones. 

 The duality transformation is performed by relaxing the linearity 
constraint on the linear superfield. Simultaneously we add to its action
chiral Lagrange multipliers that enforce the constraint. When we impose 
the field 
equations of the unconstrained superfield, we obtain the dual action of
the chiral Lagrange multiplier. 

 If we apply this transformation to nonlinear actions of linear
superfields, the corresponding duality equations are also nonlinear.
In general it is difficult to find the dual nonlinear action explicitly. 
However, for the actions with partially broken
$N=2$ supersymmetry that we described above, the presence of fermionic
superfields in the nonlinear term simplifies the problem.

\subsection{Tensor multiplet}

 We begin studying the duality transformation on the real linear
superfield. We relax the linearity constraint and we add a Lagrange 
multiplier that in this case is necessarily real

\beqs
 S & = & \int d^4 x \, d^4 \th \( -{1 \over 2} G^2 + G(\phi +\bar{\phi})
                     +2 \Psi^2 \bar{\Psi}^2 \,f \) \nn  \\
   & = & \int d^4 x \, d^4 \th \( \phi \bar{\phi} -{1 \over 2} 
       (G- \phi -\bar{\phi})^2 + 2 \Psi^2 \bar{\Psi}^2 \,f \)  
\label{phiactinG}
\eeqs

\ni
Note that once we have relaxed the linearity constraint on $G$, there 
is large class of actions which are equivalent modulo terms containing 
$D^2 G$ and $\bar{D}^2 G$. They all reduce to the same nonlinear action 
when we eliminate the Lagrange multiplier. Such terms can be reabsorbed 
in to a redefinition of the chiral Lagrange multiplier

\be
 G(\phi +\bar{\phi}) + F(G) D^2 G + \bar{F} (G) \bar{D}^2 G
 \longrightarrow G \( \phi + \bar{D}^2 \bar{F} + \bar{\phi} + D^2 F \) \ . 
\ee

\ni 
Later we will make use of this freedom to redefine the dual chiral
field and simplify the dual action. The duality equations that allow
us to find $G(\phi, \bar{\phi})$ are derived from 

\beqs
 { \delta S \over \delta G} =  0 \Rightarrow  \phi+\bar{\phi} & = & 
 G + \Psi^2 \bar{D}^2 G K + \bar{\Psi}^2 D^2 G \bar{K} 
\label{eq:dual2}      \\ 
& & + \: \Psi^\a \bar{\Psi}^{\da} \( (\bar{D}_{\da} D_\a G)  H - 
 (D_\a \bar{D}_{\da} G) \bar{H} \) + 
 O(\Psi^2 \bar{\Psi}, \Psi \bar{\Psi}^2) \nn \ ,
\eeqs

\ni  
where the functionals $K$ and $H$ depend on the bosonic superfields
$A$, $B$, (see (\ref{BGnot})) and

\beqs
 Q & \equiv & (D^2 G) (\bar{D}^2 G)  \nn \\
 P & \equiv & (D^\a \bar{D}^{\da} G) (\bar{D}_{\da} D_\a G)  \ .
\label{boson_var}
\eeqs

\ni
$K$ and $H$ are given explicitly in the appendix A. We have dropped terms 
with more than two fermionic fields because they do not contribute to 
the action

\be
 (G- \phi -\bar{\phi})^2 = \Psi^2 \bar{\Psi}^2 \( 2 Q K \bar{K} +
 {1 \over 2} P H \bar{H} - {1 \over 4} (A+B) \bar{H}^2 - {1 \over 4}
 (A-B) H^2 \) \ .
\ee

 In the action (\ref{phiactinG}) we have two terms with the
maximum number of fermionic fields, multiplied by a function of the
bosonic fields in (\ref{boson_var}). 

\be
 S =  \int d^4 x \, d^4 \th \( \phi \bar{\phi} + \Psi^2 \bar{\Psi}^2 \[ 2 f   
 - Q K \bar{K} - {1 \over 4} P H \bar{H} + {1 \over 8} (A+B) \bar{H}^2
 + {1 \over 8} (A-B) H^2 \] \) \ .
\label{interm_act}
\ee

 Acting with $N=1$ spinor derivatives on equation 
(\ref{eq:dual2}) we find the duality equations relating the 
fermionic field $\psi \equiv D \phi$ to $\Psi$ and its conjugate

\beqs
 \psi_\a & = & \Psi_\a \( 1 - Q K + {1 \over 2} (A+B) \bar{H} \) - \Psi_\b
 (\bar{D}^{\da} D^\b G) (D_\a \bar{D}_{\da} G) H \nn \\
& & + \bar{\Psi}^{\da} D^2 G \( (D_\a \bar{D}_{\da} G) (\bar{H} - \bar{K}) -
 (\bar{D}_{\da} D_\a G) H \) + 
 O(\Psi^2, \bar{\Psi}^2, \Psi \bar{\Psi}) \ .
\label{dual_eq_fer}
\eeqs 

 We have again dropped terms with too many fermionic fields which do
not contribute to the four fermion factor in (\ref{interm_act}). 
Using the fermionic duality equation (\ref{dual_eq_fer}) this
factor can be rewritten as the product of four dual fermionic 
fields multiplying some complicated function $1/M$ (see (\ref{real_fer_ap}))
of the bosonic superfields 
  
\be
 \Psi^2 \bar{\Psi}^2 = {\psi^2 \bar{\psi}^2 \over M(P,Q,A,B)}  \ .
\ee

 Acting once more with spinor derivatives on (\ref{dual_eq_fer}) and 
dropping all terms with fermionic fields we obtain nonlinear equations 

\beqs
 D^2 \phi & = & (D^2G) \, (1-4Qf-2Pf) + O(\Psi,\bar{\Psi})  \\
 \bar{D}_{\da} D_\a \phi & = & (\dblad \dla G) \: \( 1 + Q(H-K) + 
  {1 \over 2}(A+B) \bar{H} \) \nn \\
& & - \: (\dla \dblad G) \: \( Q(\bar{H} - \bar{K} ) + 
       {1 \over 2} (A-B)H \) + O(\Psi,\bar{\Psi})  \nn 
\eeqs                                                                          

\ni
and their corresponding complex conjugates. Squaring these equations
and taking the product of the last one and its conjugate we find 
the relation between the bosonic fields $P,Q,(A+B),(A-B)$ and the 
corresponding duals

\beqs
q & = &  D^2 \phi \bar{D}^2 \bar{\phi} \nn \\
p & = & - \pa \phi \pa \bar{\phi}   \nn \\
a+b & = & (i \pa \phi )^2  \nn \\
a-b & = & (i \pa \bar{\phi} )^2 
\eeqs

\ni
up to fermionic superfields which again do not contribute to $\psi^2
\bar{\psi}^2 / M(P,Q,A,B)$. These relations are remarkably simple for 
$p$ and $q$

\beqs
 A + P & = & a + p  \\
 q & = & Q \( 1 - 2 P f - 4 Q f \)^2 \ , \label{q-eqn}
\eeqs 

\ni
while for $a+b$ they are more complicated nonlinear equations (we omit
the obvious conjugate equation for $a-b$) 

\beqs
 a+b & = & (i \pa \phi)^2  \nn \\
 & = & (A-B) \[1 + Q (H-K) + {(A+B) \over 2} \bar{H} \]^2 +   
 (A+B) \[Q (\bar{H} - \bar{K}) + {(A-B) \over 2} H \]^2 \nn \\
 & & - \:2 \, P \[1 + Q (H-K) + {(A+B) \over 2} \bar{H} \]
       \[Q (\bar{H} - \bar{K}) + {(A-B) \over 2} H \]   \ .
\label{bosenic_eqn}
\eeqs

We can eliminate the variable $P$ using the first equation, and we
are left a system of three coupled nonlinear equations in $Q,A$ and $B$.
Finding a closed solution is a difficult task since we do not know in 
general how to invert  this change of variables.
One thing we can do is to solve these equations iteratively to
find the expansion of $Q,A,B$ to any given order in $p,q,a,b$. 

A different strategy that has proved more successful is to notice that
the system of equations can be solved exactly if we set $q=0=Q$

\beqs
A_0 & = & -{a(1-p)+ a^2-b^2 \over (1-p)^2 - (a^2-b^2) } \\
B_0 & = & {b \over \sqrt{(1-p-a) \[ (1-p)^2 - (a^2-b^2) \] } }\nn  \ .
\eeqs

\ni
We also notice from the duality equation (\ref{q-eqn}) that the bosonic 
field $Q$ is always at least linear in its dual $q$

\be
 Q = { q \over (1 - 2 (a+p-A) f - 4 f Q)^2 } \ .
\label{seed}
\ee

\smallskip
\ni
Assuming that the solutions $A(q,p,a,b)$, $B(q,p,a,b)$ and $Q$ have a Taylor
series expansion in $q$

\beqs
 A & = & \sum_{i=0}^{+\infty} q^i A_i(p,a,b) \nn \\
 B & = & \sum_{i=0}^{+\infty} q^i B_i(p,a,b) \nn \\
 Q & = & \sum_{i=1}^{+\infty} q^i Q_i(p,a,b) \ ,
\label{2seed}
\eeqs

\ni
we may solve for $Q$ iteratively 

\beqs
 Q_1 & = & { q \over (1 - 2 (a+p-A_0) f_0)^2 }  \ , \\
 . & & \nn \\
 . & & \nn  \\
 . & & \nn
\eeqs 

\ni
where $f_0 = f(A_0, B_0)$. We can substitute the Taylor series expansion
of $Q$, $A$, and $B$ on the {\em r.h.s.} of (\ref{bosenic_eqn}). Since the 
{\em l.h.s.} is $q$-independent, each order $n$ in $q$ on the {\em r.h.s.} 
gives equations for $A_{i \leq n}, B_{j \leq n}$ that can be 
solved\footnote{In the appendix B we
explain how this is done in some detail, when we solve similar duality 
equations for the complex linear multiplet.}. 

Finally we replace the Taylor expansions $Q=\sum_{i=0}^{+\infty} 
q^i Q_i(p,a,b)$, $A=\sum_{i=0}^{+\infty} q^i A_i(p,a,b)$, and 
$B = \sum_{i=0}^{+\infty} q^i B_i(p,a,b)$ in the action 
(\ref{interm_act}) to find

\be
 S_{dual} = \int d^4 x \, d^4 \th \( \phi \bar{\phi} + \psi^2 \bar{\psi}^2 
 \[ {\cal L}_0(p,a,b) + q {\cal L}_1(p,a,b) + 
 q^2 {\cal L}_2(p,a,b) + \dots \] \) \ .
\label{dual_act}
\ee

\ni
We are able to derive the dual action as a Taylor series in the
auxiliary superfield $q=(D^2 \phi) (\bar{D}^2 \bar{\phi})$. 
Its coefficients are analytic functions of the dual bosonic 
field $p$ and the real combination $a^2-b^2$. The $q$-independent part
of the action exactly agrees with the form proposed in \cite{RT}

\be
 S_{dual} (q=0) = \int d^4 x \, d^4 \th \( \phi \bar{\phi} + 
 { 2 \psi^2 \bar{\psi}^2 \over 1 + \pa \phi \pa \bar{\phi} + 
 \sqrt{(1+ \pa \phi \pa \bar{\phi})^2 - (\pa \phi)^2 (\pa \bar{\phi})^2 } }\) 
 \ .
\label{phi_act}
\ee

\ni 
The coefficient linear in $q$ is also remarkably simple 

\be
 {\cal L}_1 = { 2 \over (1-p+r)^2} {(1+r)^2 \over 2 r^2+r-1+p} \ , 
\ee

\ni
where $r$ is the square root appearing in ${\cal L}_0$

\be
 r=\sqrt{(1-p)^2-(a^2-b^2)} =
 \sqrt{(1+ \pa \phi \pa \bar{\phi})^2 - (\pa \phi)^2 (\pa \bar{\phi})^2 } \ .
\ee

 Higher order coefficients become increasingly complicated, but the key
feature is that they remain nonsingular around the zero {\em v.e.v.}
of the bosonic superfields. That will prove crucial when we 
perform a field redefinition of $\phi$ to eliminate the $q$ dependent
terms in (\ref{dual_act}). It is natural to expect that their presence
in the dual $N=1$ nonlinear action is somewhat arbitrary, because they
are completely irrelevant in the computation of the physical x-space 
bosonic action. 
They only contribute auxiliary field terms that are always quadratic or higher
in the auxiliary field, and they can therefore be set to zero as a 
solution to their algebraic field equations.

\subsection{Complex linear multiplet}

 We have found a nonlinear $N=1$ action of a chiral field by dualizing
the real linear
superfield. It has an on-shell nonlinear $N=2$ supersymmetry that
can be derived by defining compensating transformation laws for $\phi$
in (\ref{phiactinG}): The variation of $\phi$ and its conjugate must cancel 
the additional terms appearing in $\d S$ when we relax 
the constraint on the tensor superfield.

 Now we want to derive the nonlinear action of a chiral $N=1$ multiplet 
dual to the complex linear superfield of (\ref{Sigma_act}). We
follow a very similar argument, though the algebra and the equations
are somewhat simplified by the fact that we have kept some of the
$\S$-dependence implicit in the function $Y$. As before, we relax
the constraint on the linear superfield and add complex Lagrange 
multipliers to (\ref{Sigma_act})

\beqs
 S & = & \int d^4 x \, d^4 \th \( - \S \bar{\S} + \S \phi + \bar{\S} \bar{\phi} +  
 \tp^2 \bar{\tp}^2 {2 Y + 1 \over Y^3} \)   \nn  \\
 & = & \int d^4 x \, d^4 \th \( \phi \bar{\phi} - (\S - \bar{\phi}) 
 (\bar{\S}- \phi) + \tp^2 \bar{\tp}^2 {2 Y + 1 \over Y^3} \)  \ .
\label{Sigma_int_act}
\eeqs 

\ni
The basic duality relations are now
 
\beqs
 { \delta S \over \delta \S} = 0 \Rightarrow 
 \phi & = & \bar{\S} + \tp^2 \bar{D}^2 \S {2 \over Y^2} + \tp^2 
 D^2 \bar{\S} { \dbdfs (D^\a \bar{D}^{\da} \S)
 (\bar{D}_{\da} D_\a \bar{\S}) \over 2 Y^6} \\
 & & + \tp^\a \bar{\tp}^{\da} \[ { (\bar{D}_{\da} D_\a \bar{\S}) \over Y^2}   
 + (D_\a \bar{D}_{\da} \S) {\dbdfs D^2 \bar{\S} \bar{D}^2 \S \over 2 Y^6} \]
 + O(\tp^2 \bar{\tp}, \tp \bar{\tp}^2)   \nn   \ ,
\label{Sigma_dual}
\eeqs

where we have used (\ref{implicit}) implicitly. To simplify the notation 
we define bosonic superfields analogous to those in (\ref{boson_var})

\beqs
 \tQ & = & (D^2 \bar{\S}) (\bar{D}^2 \S) \nn \\
 \tP & = & (D^\a \bar{D}^{\da} \S)(\bar{D}_{\da} D_\a \bar{\S}) \nn \\
 {\cal A} & = & {1 \over 2} \( \ddbfs + \dbdfs \) \nn \\
 {\cal B} & = & {1 \over 2} \( \ddbfs - \dbdfs \) \ ,
\eeqs

\ni
although in fact the bosonic fields ${\cal A}$ and ${\cal B}$ always 
appear in the real combination  ${\cal A}^2 - {\cal B}^2 = 4 (Y^4+Y^3)$.
The presence of fermionic superfields in the duality equations
simplifies again our calculation of the Lagrange multiplier term 

\be
 (\S - \bar{\phi}) (\bar{\S}- \phi) = \tp^2 \bar{\tp}^2 \( 
 4 \tQ {2 Y +1 \over Y^5}  + {\tP \over Y^4} + \tQ \tP
 (\tQ + \tP) {Y+1 \over Y^9} \) \ .
\ee

\ni
Combining the last two terms in (\ref{Sigma_int_act}) we obtain

\be
 S = \int d^4 x \, d^4 \th \( \phi \bar{\phi} + \tp^2 \bar{\tp}^2 
 \[ {2 Y + 1 \over Y^3} \(1 - {4 \tQ \over Y^2} \) - {\tP \over Y^4} 
 - \tQ \tP (\tQ + \tP) {Y+1 \over Y^9} \] \) \ .
\label{O4_act_relax}
\ee

 The relevant fermionic fields are again linear in their duals, as follows
from acting with spinor derivatives on (\ref{Sigma_dual}) and its
complex conjugate

\beqs
 \psi_\b \equiv D_\b \phi & = & \( 1 - \tQ {3 Y +1 \over Y^3} \) \tp_\b - 
 {1 \over Y^2} (\bar{D}^{\da} D^\a \bar{\S}) ( D_\b \bar{D}_{\da} \S) 
 \tp_\a                                \label{Sigma_fer} \\ \nn \\
& & - D^2 \bar{\S} \( \dbdfs {\tQ +\tP \over 2 Y^6} 
 (D_\b \bar{D}_{\da} \S) + {1 \over Y^2} (\bar{D}_{\da} D_\b \bar{\S}) \)
 \bar{\tp}^{\da} + O(\tp^2, \bar{\tp}^2, \tp \bar{\tp})  \nn \ .
\eeqs

\ni
We can rewrite the four fermion numerator as before

\be
 \tp^2 \bar{\tp}^2 = \psi^2 \bar{\psi}^2 {1 \over N(\tQ,\tP,Y)} \ ,
\label{complex_fer}
\ee

\ni
where the function $N$ is defined in the appendix A (\ref{complex_fer_ap}).
Acting once more with spinor derivatives on (\ref{Sigma_fer}) 

\beqs
 D^2 \phi & = & D^2 \bar{\S} \( 1 - \tP {2Y+1 \over Y^3} - 
                  2 \tQ {2Y+1 \over Y^3} \) + O(\tp, \bar{\tp})   \\ 
 \bar{D}_{\da} D_\a \phi & = & (\bar{D}_{\da} D_\a \bar{\S}) 
   \( 1 - \tQ {2Y+1 \over Y^3} \) -  (D_\a \bar{D}_{\da} \S)   
   { (\bar{D} D \bar{\S})^2 \over 2 Y^2} 
   \( 1 - \tQ {\tQ + \tP \over Y^4} \) + O(\tp, \bar{\tp})    \nn 
\eeqs                               

\ni
we square these equations and compute the product of the last one and
its conjugate to find the duality equations relating the bosonic 
superfields $\tQ, \tP$ and $Y$ to the duals $q, p$ and $a^2-b^2$.
The relations we find are very different because of the complex nature
of the linear superfield, but the equation defining the auxiliary 
superfield $q$ has the same structure

\be 
 q = \tQ \( 1 - \tP {2Y+1 \over Y^3} - 2 \tQ {2Y+1 \over Y^3} \)^2 \ .
\ee

 This is important because we may once again solve for $\tQ = 
\tQ (q)$ iteratively and apply the same procedure as in the
$O(2)$ multiplet case to solve the nonlinear equations of 
$\tP (q,p,a^2-b^2)$ and $Y(q,p,a^2-b^2)$ order by order in $q$ (see 
appendix B). Substituting this solution on the action (\ref{O4_act_relax})
we find once more its Taylor series expansion in $q$

\beqs
 S_{dual} & = & \int d^4 x \, d^4 \th \( \phi \bar{\phi} + { 2 \psi^2 \bar{\psi}^2
 \over 1 + \pa \phi \pa \bar{\phi} + \sqrt{(1+ \pa \phi \pa
 \bar{\phi})^2 - (\pa \phi)^2 (\pa \bar{\phi})^2 } } \right. \\
& & \left. \qquad \qquad + \: q \:{2 \psi^2 \bar{\psi}^2 \over (1-p+r)^2}
  {p^2 +p-2 - 2(1-p) r^2 + (3p-4) r \over 3(1-p)^2+(4p-5)r^2-2(1-p)r} 
 + O(q^2) \)   \nn \ .
\eeqs

 The fact that the $q$-independent part of the action is the same as
in the tensor multiplet case does not come as a surprise. Both the 
$O(2)$ multiplet and the $O(4)$ multiplet free action are dual to the same 
free action of two chiral fields that realizes $N=2$ supersymmetry
on-shell. The disagreement on the $q$-dependent terms is surprising
and might  appear disturbing; however, as we see below, the $q$-dependent
terms are irrelevant since they can be removed by a field
redefinition. This is the subject of our next section.

\section{Redefinition of the dual Chiral field }

 We have found the nonlinear actions of a chiral multiplet dual to a
tensor multiplet and a complex linear multiplet. They are both of the
form 

\be
 S_{dual} = \int d^4 x \, d^4 \th \( \phi \bar{\phi} + \psi^2 \bar{\psi}^2 
 \[ {\cal L}_0 + q {\cal L}_1 + q^2 {\cal L}_2 + \dots \] \) \ .
\label{chiral_d_act}
\ee   

 Can we find a field redefinition of the dual chiral field
that preserves the first two terms in this action while we change
the rest? A suitable redefinition is the following

\be
 \phi \equiv \varphi + \bar{D}^2 \( \hps^2 \bar{\hps}^2 \,
        (D^2 \varphi)\, h(\hq, \hp, \hab) \) \ ,
\label{field_red}
\ee

\ni
where $\varphi$ is a new chiral field and we define the superfields

\beqs
 \hps_\a & \equiv & D_\a \varphi \nn \\
 \hq & \equiv &  D^2 \varphi \bar{D}^2 \bar{\varphi} \nn \\
 \hp & \equiv & - \pa \varphi \pa \bar{\varphi}   \nn \\
 \hab \equiv \hat{a}^2- \hat{b}^2 & \equiv & ( \pa \varphi )^2 
                    (\pa \bar{\varphi} )^2 \ .
\eeqs

 The real function $h$ is assumed to have a Taylor series expansion
in $\hq$ with coefficients which are analytic functions of $\hp$ and
$\hab$ around their vanishing {\em v.e.v.}\footnote{This guarantees there
exists a nonsingular inverse redefinition.}

\be
 h(\hq, \hp, \hab) = \sum_{i=0}^{+\infty} \hq^i h_i(\hp,\hab) \ .
\label{h_series}
\ee 

\ni
We must now rewrite the action (\ref{chiral_d_act}) in terms of the
new chiral field. The nice property of this
peculiar field redefinition is that it involves fermionic superfields 
and we can therefore apply the same simplifying arguments that have
proved so useful throughout. The kinetic term transforms into

\be
 \phi \bar{\phi} = \varphi \bar{\varphi} + \hps^2 \bar{\hps}^2 \hq
                  \[ 2 h + h^2 \( \hq (\hq + p) + {\hab \over 4} \) \]  \ .
\label{phiphib}
\ee

\ni
Note that the transformation of the kinetic term does not introduce
unwanted $\hq$-independent terms. As before, when we compute $\psi$ 
in terms of $\hps$ we only keep terms which are linear in the 
redefined fermionic field

\beqs
 \psi_\a & \equiv & (1 + \hq^2 h) \: \hps_\a + \hq \: h \:
 (\bar{D}^{\db} D^\b \varphi) (D_\a \bar{D}_{\db} \bar{\varphi}) \: \hps_\b \\
& & + D^2 \varphi \: h \( \hq (\bar{D}_{\da} D_\a \varphi) 
 - {1 \over 2} (\pa \varphi)^2 (D_\a \bar{D}_{\da} \bar{\varphi}) \)
 \bar{\hps}^{\da} + O(\hps^2, \bar{\hps}^2,\hps \bar{\hps}) \nn \ ,
\eeqs

\ni
because we are only interested in computing 

\be
 \psi^2 \bar{\psi}^2 = \hps^2 \bar{\hps}^2 
 \(1 + \hq {\cal M} (h, \hq, \hp, \hab) \) \ .
\label{redef_fer}
\ee

\ni
Here the function ${\cal M}$ is a finite polynomial on the real variables
$h, \hq,\hp$ and $\hab$ (see appendix A (\ref{redef_fer_ap})). Acting
with an additional spinor derivative on $\psi$ it is easy to see that 

\be
 \dblad \dla \phi = \dblad \dla \varphi  \ .
\label{abpsame}
\ee

\ni
Therefore the bosonic superfields $(a^2-b^2)$, and $p$ are the same as their
redefined partners $\hab$ and $\hp$ up to irrelevant terms containing
fermionic fields. This guarantees that the piece of the action without
auxiliary superfields remains unchanged 

\be
 \hps^2 \bar{\hps}^2 {\cal L}_0 (p, a^2-b^2) = \hps^2 \bar{\hps}^2 
 {\cal L}_0 (\hp, \hab) \ .
\ee

\ni 
On the other hand the auxiliary bosonic superfield transforms nontrivially

\be
 D^2 \phi =  D^2 \varphi \[ 1+h \( \hq (\hq + p) + {\hab \over 4}\) \]
 + O(\hps, \bar{\hps}) \ .
\ee

\ni
As a result we obtain the following nonlinear relation

\be
 q =  \hq \[ 1+h \( \hq (\hq + p) + {\hab \over 4}\) \]^2 \ .   
\label{qvarphieq}     
\ee  

\ni
Collecting all these results the redefined dual action may be written
as 

\beqs
 S_{dual} & = & \int d^4 x \, d^4 \th \( \varphi \bar{\varphi} + \hps^2 
 \bar{\hps}^2 \,\hq \[2 h + h^2 \( \hq (\hq + p) + {\hab \over 4} \) \] 
 \right. \\
 & & + \left. \hps^2 \bar{\hps}^2 \; \(1 + \hq {\cal M} (h,\hq, \hp, \hab) \)
 \[ {\cal L}_0 (\hp,\hab) + \hq {\cal L}_1 (\hp,\hab) + 
 \hq^2 {\cal L}_2 (\hp,\hab) + \dots \]   \) \nn \ .
\label{varphi_act}             
\eeqs

Thus we have shown that is possible to make the dependence of the 
dual action on auxiliary superfields $\hq$ completely arbitrary, while 
the physical part of the action remains unchanged. In particular we 
may now replace the function $h$ by its Taylor
series expansion (\ref{h_series}) in the auxiliary superfield $\hq$
and solve for the coefficients $h_i$ that eliminate the $\hq$ dependence of
the action at each order. We begin imposing the condition that the 
linear piece of the redefined action vanishes

\be
 0 = h_0^2 {\hab \over 4} \(1 + \hp {\cal L}_0 + {\hab \over 4} 
 {\cal L}_1 \) + 2 h_0 \(1 + \hp {\cal L}_0 + {\hab \over 4} {\cal L}_1 \) 
 + {\cal L}_1  \ .
\ee  

\ni
This gives a quadratic equation for the coefficient $h_0$. To have an 
invertible field redefinition we choose the nonsingular 
root

\be
 h_0 =  - {4 \over \hab} \( 1 - \sqrt{1- {\hab {\cal L}_1 \over  
 4 (1 + 4 \hp {\cal L}_0 + \hab {\cal L}_1 ) } } \) \ .
\ee

 Next we impose that the term quadratic in $\hq$ vanishes. The result
is a linear equation on $h_1$, with coefficients depending on 
$h_0,{\cal L}_0,{\cal L}_1,{\cal L}_2$ and the bosonic superfields 
$\hp,\hab$. We can continue this process and we find at every order
$\hq^n$ a linear equation on $h_{n-1}$ with the same linear coefficient

\be 
 0 = 2 h_n \( (1 + h_0 {\hab \over 4}) (1 + \hp {\cal L}_0) + {\hab \over
 4} {\cal L}_1 \) + C_n (h_{i<n},{\cal L}_{j<n+1},\hp,\hab) \ .
\ee

 Thus we can eliminate the auxiliary superfield dependence at each
order in the Taylor series expansion of the action. The last
consistency check we must perform is to make sure that the solution to
each equation 

\be
 2 h_n = - {C_n \over (1 + h_0 {\hab \over 4}) (1 + \hp {\cal L}_0) 
 + {\hab \over 4} {\cal L}_1 }
\ee

\ni
is nonsingular. This is automatically guaranteed because the linear 
coefficient of all these equations has a non-vanishing {\em v.e.v.}, 
and $C_n$ is a function of nonsingular quantities.

\section{Nonlinear Bosonic action in x-space}

 In this section we integrate the Grassmann coordinates on the 
nonlinear action (\ref{Sigma_act}) of the complex linear superfield.
We only analyze the bosonic part; this reproduces the
dependence of the 3-brane transverse coordinates on the longitudinal 
ones. The
supersymmetric fermionic partner coordinates can be trivially derived
from the additional terms. Since the duality between the complex
linear superfield and its dual chiral superfield does not exchange the
physical x-space components, the x-space action of the bosonic fields
must be the same as that obtained from (\ref{phi_act}). Historically 
the supersymmetric action (\ref{phi_act}) was actually guessed \cite{RT}
(see also \cite{BG}) from the bosonic one 

\be
 S_{bos} = \int d^4 x \sqrt{ 1 + 2 (\pa \phi_o \pa \bar{\phi}_o) + 
 (\pa \phi_o \pa \bar{\phi}_o)^2 - (\pa \phi_o)^2 (\pa \bar{\phi}_o)^2} \ ,
\label{bos_p_act}
\ee  

\ni
where the field $\phi_o = X_4+iX_5$ denotes the physical scalar in the
chiral superfield $\phi$ and we integrate over the longitudinal
coordinates $d^4 x = dx_0 dx_1 dx_2 dx_3$.  

 Before we start our calculation let us define a useful notation for
the components of the complex linear superfield

\beqs
 \s & \equiv & \S |_{\th=0}  \nn \\
 \chi_\a & \equiv & D_\a \S |_{\th=0}  \nn  \\
 \lambda_{\da} & \equiv & \bar{D}_{\da} \S |_{\th=0}  \nn   \\
 F_{\a \da} & \equiv & D_\a \bar{D}_{\da} \S |_{\th=0} 
\eeqs

 Since we are studying only the bosonic part of the x-space action,
when we integrate the Grassmann coordinates in (\ref{Sigma_act}) we
drop all terms containing the fermionic fields $\chi_\a$,
$\lambda_{\da}$ and their conjugates. In addition, the auxiliary field
$D^2 \S |_{\th=0}$ and its conjugate always enter quadratically and
they can be set to zero using their algebraic field equations. Thus
the bosonic action is

\be
 S = \int d^4 x \( i F \cdot \pa \bar{\s} + i \bar{F} \cdot \pa \s - 
 F \cdot \bar{F} + {1 \over 2} \pa \s \cdot \pa \bar{\s} + 
 {1 \over 4} F^2 \bar{F}^2 {2 y + 1 \over y^3} \) \ ,
\label{bos_s_act}
\ee

\ni
where the field $y$ represents now its lowest component of the superfield
$Y$. It is defined by the identity $4 (y^4 +y^3)
= F^2 \bar{F}^2$. The auxiliary field $F_{\a \da}$ does not enter
quadratically in the bosonic action, but it is not dynamical and it
may be eliminated imposing its algebraic field equation 

\be
 F_{\a \da} = i \pa_{\a \da} \s + \bar{F}_{\a \da} {F^2 \over 2 y^2} \ ,
\ee

\ni
and substituting $\bar{F}_{\a \da}$ by the corresponding conjugate 
field equation. The solution is 

\be
 F_{\a \da} =  i (\pa_{\a \da} \s) y +  i (\pa_{\a \da} \bar{\s}) 
 {F^2 \over 2 y^2} \ .
\ee

 Squaring this relation and the corresponding conjugate we find after
some algebraic manipulations

\beqs
 (\pa \s)^2 \bar{F}^2 & = & (\pa \bar{\s})^2 F^2 = -
 (1 + \pa \s \cdot \pa \bar{\s}) {4 (y^3+y^2) \over 2 y + 1}  \nn \\
 y & = & {1 \over 2} \( 1 + {1 + \pa \s \cdot \pa \bar{\s} \over 
            \sqrt{ (1 + \pa \s \cdot \pa \bar{\s})^2 - 
                           (\pa \s)^2 (\pa \bar{\s})^2 } } \) \ .
\eeqs

 Finally we substitute these solutions in the action (\ref{bos_s_act})
we find the same type of 3-brane action as in (\ref{bos_p_act})

\be
 S_{bos} = \int d^4 x \sqrt{ 1 + 2 (\pa \s \pa \bar{\s}) + 
 (\pa \s \pa \bar{\s})^2 - (\pa \s)^2 (\pa \bar{\s})^2} \ .
\ee

 This exactly the bosonic action we expect in agreement with the fact
that the physical fields $\phi_o = \bar{\s}$ are not affected by the $N=1$
duality transformation between the superfields $\phi$ and $\bar{\S}$.

\section{Superpotential deformation on the $O(4)$ nonlinear action}

 In this last section we study the effect of adding a superpotential 
deformation (\ref{deform}) to the complex linear multiplet action
(\ref{O4_nonl_act}) with partially broken supersymmetry

\be
S_{O(4)} = \int d^4 x \, d^4 \th \( \Phi \bar{\Phi} - \S \bar{\S} + 
 {1 \over 2} X^2 + \b X(\Phi + \bar{\Phi}) + {\b \over 2} 
 (\S^2 + \bar{\S}^2) \) \ .
\label{super_def}
\ee

 It is possible to replace the constrained chiral field $\Phi$ by its   
solution (\ref{const_soll}) and obtain the equation of motion for the 
unconstrained superfield $X$. The bosonic x-space action can then be
computed as in the previous section. The counting of
fermionic fields again simplifies our calculations, but the implicit
dependence of $D^2 X$ on the scalar component of ${\cal P}$ and its
conjugate cannot be solved exactly for a generic value of the
parameter $\b$. For the special values $\b=\pm 1$ there is a solution 
but the x-space action vanishes. 

 A more illuminating analysis is provided by the dualization of the
action (\ref{super_def}) with partially broken supersymmetry to a 
nonlinear action of a chiral Lagrange multiplier. Our task is slightly
simplified if we redefine the auxiliary superfield in (\ref{super_def}) 
by a shift

\be
X \longrightarrow {\cal X} = X + \b (\Phi + \bar{\Phi}) \ .
\ee

\ni
 The action (\ref{super_def}) can be written

\be 
 S_{O(4)} = \int d^4 x \, d^4 \th \( (1 - \b^2) \Phi \bar{\Phi} + {1 \over 2} 
 {\cal X}^2 - \S \bar{\S} + {\b \over 2} (\S^2 + \bar{\S}^2) \) 
\label{redef_acti}
\ee
 
\ni
where the nilpotent antichiral field (\ref{const_soll}) can be expressed 
in terms of this redefined ${\cal X}$

\be
 \bar{\Phi} = - {1 \over 2} {\tp^a \tp_\a \over 1 + D^2 {\cal X}    
 - \b D^2 \Phi } \ .
\label{D2phi}
\ee

 Counting the number of fermionic superfields in $\Phi \bar{\Phi}$
we realize once more that only the bosonic part of $(D^2 \Phi)$ and
$\bar{D}^2 \bar{\Phi}$ contributes to its denominator. We can therefore
solve for this superfields after acting with spinor derivatives on 
(\ref{D2phi}) and its complex conjugate 

\beqs
 (D^2 \Phi)_{bos} & = & {1 \over 2} \: { {\cal A} + {\cal B} \over 1 + 
 \bar{D}^2 {\cal X} - \b (\bar{D}^2 \bar{\Phi})_{bos} } \nn  \\
 (\bar{D}^2 \bar{\Phi})_{bos} & = & {1 \over 2} \: { {\cal A} - {\cal B}
 \over 1 + D^2 {\cal X} - \b (D^2 \Phi)_{bos} }  \ .
\eeqs
 
We substitute the solution of this system of quadratic equations in the
action (\ref{redef_acti}) and we find

\be
 S_{O(4)} = \int d^4 x \, d^4 \th \( (1 - \b^2) {\cal F}  
 \tp^\a  \tp_\a \bar{\tp}^{\da} \bar{\tp}_{\da} + {1 \over 2} {\cal X}^2
  - \S \bar{\S} + {\b \over 2} (\S^2 + \bar{\S}^2) \) \ ,
\ee

\ni
where 

\be
 {\cal F} = {1 \over 2} \( {\cal Y} \bar{\cal Y} - \b {\cal A} + 
 \sqrt{ ({\cal Y} \bar{\cal Y})^2 - 2 \b {\cal A} {\cal Y} \bar{\cal Y} + 
 \b^2 {\cal B}^2 } \)^{-1} \ ,\;\;\; {\cal Y} = 1 +
  D^2 {\cal X} \ .
\ee

 The field equation of the unconstrained superfield ${\cal X}$ reveals
that it is at least quadratic in $\tp$ and in $\bar{\tp}$ 

\be
 {\cal X} = - \(\tp^\a  \tp_\a \bar{\cal Y} ({\cal A} - {\cal B}) + 
 \bar{\tp}^{\da} \bar{\tp}_{\da} {\cal Y} ({\cal A} + {\cal B})  \)
 { (1 - \b^2) {\cal F} \over \sqrt{ ({\cal Y} \bar{\cal Y})^2 - 2 \b    
 {\cal A} {\cal Y} \bar{\cal Y} + \b^2 {\cal B}^2 } } + 
 O(\tp^2 \bar{\tp}, \tp \bar{\tp}^2) \ .
\label{red_aux_eqn}
\ee

\ni
Therefore ${\cal X}^2$ contains again the maximal number of fermionic
superfields and only the bosonic part of ${\cal Y}$ contributes to the
action. Acting with spinor derivatives on (\ref{red_aux_eqn}) and
keeping only terms without fermionic superfields we find as in the
$\b=0$ case that $(\bar{\cal Y})_{bos} = ({\cal Y})_{bos}$. The
algebraic equation defining $({\cal Y})_{bos}$ in terms of ${\cal A}$
and ${\cal B}$ is now a higher order order polynomial that we
cannot solve exactly (to simplify the notation we drop from now on the 
bosonic subscript)

\be
 {\cal Y} -1 = {(1 - \b^2) \({\cal A}^2 - {\cal B}^2 \) {\cal Y} {\cal F}
 \over \sqrt{ ({\cal Y}^4 - 2 \b {\cal A} {\cal Y}^2 
 + \b^2 {\cal B}^2 } } \ . 
\ee

 Keeping the dependence of the action on ${\cal Y}$ 
explicit does not help us solve the duality equations in this case. We
may however solve for ${\cal Y}= {\cal Y} ({\cal A},{\cal B})$
iteratively and that in turn allows us to solve the duality
equations also iteratively.

 Substituting ${\cal X}^2$ by its field equation in the action, we 
introduce Lagrange multipliers that enforce the
linearity constraint on a relaxed $\S$ 

\beqs
 S & = & \int d^4 x \, d^4 \th \(  \tp^\a  \tp_\a \bar{\tp}^{\da} 
 \bar{\tp}_{\da}
 (1 - \b^2) {\cal F} \( 1 + {(1 - \b^2) {\cal F} 
 \({\cal A}^2 - {\cal B}^2 \) {\cal Y}^2  \over 
 {\cal Y}^4 - 2 \b {\cal A} {\cal Y}^2 + \b^2 {\cal B}^2 } \) \right.  \nn \\
& & \left. \qquad \qquad - \S \bar{\S} + {\b \over 2} 
 (\S^2 + \bar{\S}^2) + \S \phi + \bar{\S} \bar{\phi} \) \ .
\eeqs

 To manipulate the duality equations that we obtain from the
functional differentiation with respect to the unconstrained $\S$ and
$\bar{\S}$, rewrite the last line in matrix form

\be
S = S_o + \int d^4 x \, d^4 \th {1 \over 2} \[ (\S, \bar{\S} + (\phi, \bar{\phi})
 {\bf M}^{-1} \] {\bf M} \[ \( \begin{array}{c} \S \\ 
                                              \bar{\S} \end{array} \)
 + {\bf M}^{-1} \( \begin{array}{c} \phi \\ \bar{\phi} \end{array} \) \]
 - { 1 \over 2} (\phi, \bar{\phi}) {\bf M}^{-1} 
 \( \begin{array}{c} \phi \\ \bar{\phi} \end{array} \) \ ,
\ee

\ni
where 

\be
 {\bf M} = \( \begin{array}{cc}  \b & -1 \\ -1 & \b \end{array} \) \ .
\ee

 Obtaining the duality equations is now a straightforward calculation very
similar to what we did before

\be
 {\bf M} \[ \( \begin{array}{c} \S \\ \bar{\S} \end{array} \)
 + {\bf M}^{-1} \( \begin{array}{c} \phi \\ \bar{\phi} \end{array} \) \]
 = \( \begin{array}{c} {\d S_o \over \d \S} \\ \\ {\d S_o \over \d \bar{\S}} 
       \end{array} \) \ .
\ee 

 The fields ${\d S_o \over \d \S}$ and its conjugate are again quadratic in 
fermionic superfields and the analysis resembles closely that of the
$\b=0$ case. Since we know that the dependence on the auxiliary
superfield ${\cal Q}$ decouples when we go to x-space components and
we are mostly interested in finding out if the superpotential
deformation introduces any new qualitative features, we will simplify
our calculation by setting $q=0={\cal Q}$ in the duality equations.

 We follow the same steps as in the $\b=0$ case and we find similar
duality equations with a more complicated dependence on the
superfields ${\cal Y}, {\cal A}, {\cal B}, {\cal P}$.
In this case it is not possible to solve the equations exactly even after
setting ${\cal Q}=0=q$. However, the bosonic variables $a,b,p$ are at
least linear on their duals ${\cal P}, {\cal A}, {\cal B}$ and we can
therefore solve the duality equations iteratively to find the Taylor
series expansion of ${\cal P}(a,b,p), {\cal A}(a,b,p), {\cal B}(a,b,p)$.

 Expanding the action to a given order $n$ on ${\cal P}, {\cal A}, 
{\cal B}$ and substituting the iterative solution up to that order 
we obtain the expansion of the dual action to $n$-th order in $a,b,p$.
What we find after this tedious but straightforward calculation is
that the dual chiral field gets rescaled and the action is
multiplied by an overall constant

\be
 S_{dual}(q=0) = (1 - \b^2) \int d^4 x \, d^4 \th \( 
 {\phi \bar{\phi} \over (1-\beta^2)^2} + {\psi^2 \bar{\psi}^2 \over 
 (1- \b^2)^4} {\cal L}_0 \( {p \over (1 - \b^2)^2}, 
 {a^2 - b^2 \over (1 - \b^2)^4 } \) \) 
\ee

 The rescaling of the dual chiral field becomes singular when 
$\b= \pm 1$. For other values of the deformation parameter $\b$ the 
superpotential does not seem to add any significant new feature (see also
\cite{deo_gates}).

\section*{\Large \bf Acknowledgments}

This research was in part supported by NSF grant No Phy 9722101.

\section*{\LARGE\bf Appendix}

\appendix

\section{Explicit formulas}

In this appendix we include the explicit form of various expressions
appearing in sections 3 and 4. The basic duality duality equation
(\ref{eq:dual2}) is defined by 

\beqs
 \phi+\bar{\phi} & = & G + \Psi^2 (\bar{D}^2 G) K + 
 \bar{\Psi}^2 (D^2 G) \bar{K} + \Psi^\a \bar{\Psi}^{\da} 
 \( (\bar{D}_{\da} D_\a G)  H - (D_\a \bar{D}_{\da} G) \bar{H} \)   \nn  \\
 \nn \\
K & = & 4 f + 2 (A + B) (f_A + f_B) - 2 P (f_A - f_B) \nn \\
H & = & 2f + 2 Q (f_A - f_B) + (A-B) (f_A + f_B) \ .
\eeqs

\ni
The product of four fermionic superfields derived from
(\ref{dual_eq_fer}) is

\beqs
 \psi^2 \bar{\psi}^2 & = & \Psi^2 \bar{\Psi}^2 M  \nn \\
 M & = & g_1 \, \bar{g}_1 + Q^{2} \: g_2 \, \bar{g}_2     
   + Q P (g_3 \bar{g}_3 + g_4 \bar{g}_4) - Q (A+B) g_3 \bar{g}_4 
  - Q (A-B) \bar{g}_3 g_4  \\ \nn \\
& & + \: Q P \( g_3 (A+B) \bar{H}^2 + \bar{g}_3 (A-B) H^2 \) 
  - Q P \( g_4 (A-B) \bar{H}^2 + \bar{g}_4 (A+B) H^2 \)  \nn \\
& & + \: Q P \( {3 \over 4}(A^2-B^2) - P^2 \) H^2 \bar{H}^2  \ , \nn 
\label{real_fer_ap}
\eeqs       

\ni
where       

\beqs
 g_1 & = & \( 1 - Q K + {A+B \over 2} \bar{H} \)^2 + 
      {A^2-B^2 \over 4} H^2 - P \( 1 - Q K + {A+B \over 2} \bar{H} \) H 
           \nn    \\
 g_2 & = & {A+B \over 2} (\bar{H} - \bar{K})^2 + {A-B \over 2} H^2 
            - P (\bar{H} - \bar{K}) H  \nn \\  
 g_3 & = & 2 (\bar{H} - \bar{K}) \( 1 - Q K + {A+B \over 2} \bar{H} \) 
           \nn     \\ 
 g_4 & = & 2 H \( 1 - Q K + {A+B \over 2} \bar{H} \) + 
           (A+B)(\bar{H} - \bar{K}) \ .
\eeqs       

\ni
Similarly, in the complex linear superfield action the product of 
four fermionic superfields (\ref{complex_fer}) is defined by

\beqs
 \psi^2 \bar{\psi}^2 & = & \tp^2 \bar{\tp}^2 N  \nn \\
 N & = & e_1^2 + \tQ^2 e_2 + {1 \over 4} \tQ \tP \(e_3^2 + e_4 \bar{e}_4 \)
       - {1 \over 4} \tQ e_3 \(\bar{e}_4 (\bar{D} D \S)^2 + 
            e_4 (D \bar{D} \bar{\S})^2 \) \\
 & & + \: {\tQ \over Y^4} e_3 \( \tP^2 - 2(Y^4+Y^3) \) - {\tQ \over 2 Y^4}
    \tP \( e_4(D \bar{D} \bar{\S})^2 + \bar{e}_4 (\bar{D} D \S)^2 \) \nn \\ 
 & & + \: \tQ {\tP \over Y^8} \( \tP^2 - 3 (Y^4+Y^3) \)  \nn     \ , 
\label{complex_fer_ap}
\eeqs

\ni 
where

\beqs
 e_1 & = & \( 1- {3Y+1 \over Y^3} \tQ \)^2 - {\tP \over Y^2} 
         \( 1- {3Y+1 \over Y^3} \tQ \) + {Y+1\over Y}    \nn \\          
 e_2 & = & {Y-1 \over Y^5} \[ 1 + {\tP \over Y^4}(\tQ + \tP) 
           + {Y-1 \over Y^5}(\tQ + \tP)^2 \]^2    \nn  \\
 e_3 & = & {2 \over Y^2} \[ {Y+1 \over Y^3} (\tQ + \tP) - 
                        \( 1- {3Y+1 \over Y^3} \tQ \)  \] \nn  \\
 e_4 & = & {(D \bar{D} \S)^2 \over Y^6} (\tQ + \tP) 
                           \( 1- {3Y+1 \over Y^3} \tQ \)    \ .     
\eeqs

\ni   
Finally, the product of four dual fermionic superfields (\ref{redef_fer})
is after the redefinition (\ref{field_red})

\beqs
 \psi^2 \bar{\psi}^2 & = & \hps^2 \bar{\hps}^2 
               \(1 + \hq {\cal M} (h, \hq, \hp, \hab) \)  \nn   \\
\(1 + \hq {\cal M} (h, \hq, \hp, \hab) \) & = & d_1 + \hq^2 d_2 + 
 {1 \over 4} \hq \hp ( d_3^2 + d_4 \bar{d}_4 ) + {1 \over 4}
 \hq d_3 \( \bar{d}_4 (\pa \varphi)^2 + d_4 (\pa \bar{\varphi})^2 \)
     \nn \\
& & + \: \hq^3 h^2 d_3 \( \hp^2 - {\hab \over 2} \) + \hq^3 \hp 
 \( \bar{d}_4 (\pa \varphi)^2 + d_4 (\pa \bar{\varphi})^2 \) \nn \\
& & \: + \hq^5 h^4 \hp \( \hp^2 - {3 \over 4} \hab \) 
\label{redef_fer_ap}
\eeqs

\ni
where

\beqs
 d_1 & = & {1 \over 4} \[ 2 (1 + h \hq^2)^2 + \hq^2 h^2 {\hab \over 2} +
     2 \hq h (1 + h \hq^2) \hp \]^2           \nn    \\
 d_2 & = & {1 \over 4} h^4 \hab \(\hq^2 + \hq \hp + {\hab \over 4}\)^2
            \nn  \\
 d_3 & = & 2 \hq h \( 1 + \hq^2 h + h {\hab \over 4}\)  \nn \\
 d_4 & = & - h (1 + \hq^2 h) (\pa \varphi)^2 
\eeqs

\section{Solution to nonlinear equations}       

We must still provide a derivation of the solution to the system of
nonlinear equations relating bosonic superfields in the complex linear
multiplet and the corresponding duals. The analysis of the tensor 
multiplet is very similar and can be quickly reproduced.

First we note that the system of equations has an exact solution when
$q = 0 = \tQ$ (we disregard the other two solutions for $\tQ$ since
we are basically expanding around the zero {\em v.e.v.} of this field)

\beqs
 Y_0 & = & {1 \over2} \( 1+ {1 - p \over \sqrt{(1-p)^2-(a^2-b^2)}} \) 
     \nn    \\
 \tP_0 & = & {4(Y_0^3+Y_0^2)+p Y_0 \over 2 Y_0 +1} \ .
\eeqs

\ni
As we mentioned before, solving $\tQ$ iteratively we find 

\be
 \tQ = {q \over \(1 - {2Y+1 \over Y^3} \tP \)^2 } +
     \sum_{n=2}^{+\infty}  q^n \tQ_n(\tP,Y)
\ee

\ni
and substituting it into our duality equations we obtain the expansion 

\beqs
 p(\tQ,\tP,Y) & = & \sum_{i=0}^{+\infty} q^i p_i(\tP,Y) \nn \\
 (\pa \phi)^2 (\pa \bar{\phi})^2 \equiv c & = & 
        \sum_{i=0}^{+\infty} q^i c_i(\tP,Y) \ .
\label{Taylor_dual} 
\eeqs

\ni
The bosonic superfields $\tP,Y$ that solve the duality equations are
functions of the dual variables $\tP=\tP(q,p,c)$, $Y=Y(q,p,c)$.
Assuming that these solutions have a Taylor series expansion in the
auxiliary superfield

\beqs
 \tP & = & \tP_0(p,c) + q \: \tP_1(p,c) + q^2 \: \tP_2(p,c) + 
           \dots        \nn    \\
 Y & = & Y_0(p,c) + q \: Y_1(p,c) + q^2 \: Y_2(p,c) \ ,
\label{assume_sol}
\eeqs   

\ni
we may replace this formal solution in (\ref{Taylor_dual})

\beqs
 p & = & p_0(\tP_0, Y_0) + q \tP_1 {\pa p_0 \over \pa \tP}{\big |}_{\tP=\tP_0} 
    + q Y_1 {\pa p_0 \over \pa Y}{\big |}_{Y=Y_0} + q p_1(\tP_0,Y_0)
    + O(q^2)            \nn                \\
 c & = & c_0(\tP_0,Y_0) + q \tP_1 {\pa c_0 \over \pa \tP}{\big |}_{\tP=\tP_0} 
    + q Y_1 {\pa c_0 \over \pa Y}{\big |}_{Y=Y_0} + q c_1(\tP_0,Y_0) + O(q^2)
\eeqs

 The variables $p$ and $c$ are independent of $q$ and therefore only
the homogeneous term on the {\em r.h.s.} can be nonzero. At linear
order in $q$ we find  

\be
 \( \begin{array}{cc} 
       {\pa p_0 \over \pa \tP} |_{\tP=\tP_0} & 
       {\pa p_0 \over \pa Y} |_{Y=Y_0} \\ \\
       {\pa c_0 \over \pa \tP} |_{\tP=\tP_0} & 
       {\pa c_0 \over \pa Y} |_{Y=Y_0}    \end{array} \)
  \( \begin{array}{c} 
         \tP_1 \\ \\  Y_1      \end{array} \) = -
  \( \begin{array}{c} 
         p_1 \\  \\c_1      \end{array} \)    \ .
\ee

Inverting the jacobian it is very straightforward to find the linear
coefficient of the solution (\ref{assume_sol}). The equations obtained from
higher orders in $q$ provide the higher coefficients.

\end{document}